\documentclass[ reprint, superscriptaddress, amsmath,amssymb,aps]{revtex4-2}
\usepackage{graphicx}
\usepackage{dcolumn}
\usepackage{bm}
\usepackage[utf8]{inputenc}
\usepackage{color}
\usepackage{soul}
\begin{document}

\preprint{APS/123-QED}

\title{Directional effects of antiferromagnetic ordering on the electronic structure in NdSb}

\author{Yevhen Kushnirenko}
\affiliation{Division of Materials Science and Engineering, Ames Laboratory, Ames, Iowa 50011, USA}
\affiliation{Department of Physics and Astronomy, Iowa State University, Ames, Iowa 50011, USA}

\author{Brinda Kuthanazhi}
\affiliation{Division of Materials Science and Engineering, Ames Laboratory, Ames, Iowa 50011, USA}
\affiliation{Department of Physics and Astronomy, Iowa State University, Ames, Iowa 50011, USA}

\author{Lin-Lin Wang}
\affiliation{Division of Materials Science and Engineering, Ames Laboratory, Ames, Iowa 50011, USA}

\author{Benjamin Schrunk}
\affiliation{Division of Materials Science and Engineering, Ames Laboratory, Ames, Iowa 50011, USA}


\author{Evan O'Leary}
\affiliation{Division of Materials Science and Engineering, Ames Laboratory, Ames, Iowa 50011, USA}
\affiliation{Department of Physics and Astronomy, Iowa State University, Ames, Iowa 50011, USA}

\author{Andrew Eaton}
\affiliation{Division of Materials Science and Engineering, Ames Laboratory, Ames, Iowa 50011, USA}
\affiliation{Department of Physics and Astronomy, Iowa State University, Ames, Iowa 50011, USA}



\author{P. C. Canfield}
\affiliation{Division of Materials Science and Engineering, Ames Laboratory, Ames, Iowa 50011, USA}
\affiliation{Department of Physics and Astronomy, Iowa State University, Ames, Iowa 50011, USA}

\author{Adam Kaminski}
\email[]{Corresponding author: kaminski@ameslab.gov}
\affiliation{Division of Materials Science and Engineering, Ames Laboratory, Ames, Iowa 50011, USA}
\affiliation{Department of Physics and Astronomy, Iowa State University, Ames, Iowa 50011, USA}

\begin{abstract}
    The recent discovery of unconventional surface state pairs, which give rise to Fermi arcs and spin textures, in antiferromagnetically ordered  NdBi raised the interest in rare-earth monopnictides. Several scenarios of antiferromagnetic order have been suggested to explain the origin of these states with some of them being consistent with the presence of non-trivial topologies. In this study, we use angle-resolved photoemission spectroscopy (ARPES) and density-functional-theory (DFT) calculations to investigate the electronic structure of NdSb. We found the presence of distinct domains that have different electronic structure at the surface. These domains correspond to different orientations of magnetic moments in the AFM state with respect to the surface. We demonstrated remarkable agreement between DFT calculations and ARPES that capture all essential changes in the band structure caused by transition to a magnetically ordered state.
\end{abstract}

\maketitle


\section{Introduction}

Many rare-earth monopnictides order antiferromagnetically at low temperatures due to moment bearing rare earth ions \cite{nereson_1971, bartholin1979hydrostatic, Paul_Phil1992, KumigashiraPRB1996, Paul_Jcryst2001, Yun_PRB2017, Kuroda_2018}.
In the last decade, some materials of this family were predicted to host topological Weyl states \cite{GuoNPJ2017, li2017predicted, ZhuPRB2020}, while, in some other materials, topological surface Dirac states are expected to be already present in the paramagnetic state \cite{zeng2015topological, DuanCommPhys2018} due to band inversion. These predictions are supported by numerous experimental studies \cite{wu2016asymmetric, niu2016presence, nayak2017multiple, lou2017evidence, SatoCeBi, Kuroda_2018, sakhya2022behavior, matt2020}. In a recent ARPES study \cite{SchrunkNature2022}, reported emergence of unconventional surface states (SS) below AFM transition temperature. These new states undergo Kaminski-Canfield (K-C) splitting  that leads to formation of Fermi arcs.

Similarly to other rare-earth monopnictides, NdSb and NdBi have a rock salt crystal structure and undergo AFM transition at low temperatures. This transition was observed in neutron diffraction  \cite{nereson1972neutron, Schobinger_1973, mukamel1985phase, manfrinetti2009magnetic}, magnetization \cite{bak1973magnetic, zhou2017field, wang2018topological}, resistivity, and specific heat \cite{wakeham2016large} measurements. The transition temperatures reported in these studies for NdSb and NdBi are 15 K and 24 K, respectively. In contrast to some other rare-earth monopnictides \cite{rossat1977phase,kuthanazhi2022magnetisation},  NdSb and NdBi have only one AFM phase with the magnetic structure shown in Fig.~1(a).
Two recent ARPES studies \cite{kushnirenko2022rare, SakhyaNdSb2022} reported changes in the electronic structure of NdSb upon AFM transition. However, there was a substantial difference in the band dispersion present below AFM transition. One study \cite{SakhyaNdSb2022} showed the development of two additional features at $\Gamma$-point, one of which forms a small round pocket on the Fermi surface (FS). In contrast another study \cite{kushnirenko2022rare} showed the development of surface state dispersion at $\sim$0.2 $\AA^{-1}$ away from the $\Gamma$-point. These surface states form arcs and elliptical pockets of the Fermi surface. Similar surface state dispersions were also observed in NdBi \cite{SchrunkNature2022}.

Here, we show the existence of different variations of band structures in NdSb below Neel temperature using laser ARPES. These variations coexist in the same sample at different surface locations. Also, we present DFT calculations of the band structure in the AFM ordered NdSb. By comparing the experimental results with DFT calculations, we show that observed variants of band structures are attributed to domains that have different directions of AFM1 ordering. In addition, we show that magnetically ordered NdBi can have band structure different from the one reported before. This indicates that NdBi can have different domains as well.

\section{Experimental details}
Single crystals of NdSb crystals were grown out of Sn flux using an initial concentration of Nd$_4$Sb$_4$Sn$_{96}$. The elements were weighed out into a fritted alumina crucible set (Canfield Crucible Set) \cite{Canfield2016Use, FrittURL} and sealed in a fused silica tube under partial pressure of argon. The prepared ampules were heated up to 1100$^\circ$~C over 4 hours and held there for 5 hours.
This was followed by a slow cooling to the decanting temperature of 800$^\circ$~C over 100 hours and decanting of the excess flux using a centrifuge \cite{Canfield_2019}. The cubic crystals obtained were stored and handled in a glovebox under Nitrogen atmosphere.

Most of the ARPES data was collected using vacuum ultraviolet (VUV) laser ARPES spectrometer that consists of a Scienta DA30 electron analyzer, picosecond Ti:Sapphire oscillator and fourth-harmonic generator \cite{jiang2014tunable}. Data from the laser based ARPES were collected with 6.7~eV photon energy. Angular resolution was set at $\sim$ 0.1$^{\circ}$ and 1$^{\circ}$, along and perpendicular to the direction of the analyser slit respectively, and the energy resolution was set at 2~meV. The VUV laser beam was set to vertical polarization. The diameter of the photon beam on the sample was $\sim 15\,\mu$m.
Supplementary He-lamp-based ARPES measurements were carried out using R8000 analyzer and GammaData helium discharge lamp with custom focusing optics. The diameter of the photon beam on the sample was $\sim1 $mm.

\section{Results and Discussion}
Fig.~1(b) shows a two-dimensional projection of bulk Fermi surface of paramagnetic NdSb calculated using density functional theory (DFT). As well as other rare-earth monopnictides\cite{Yun_PRB2017, Kuroda_2018, DuanCommPhys2018, SatoCeBi, SchrunkNature2022, kushnirenko2022rare}, it has several hole pockets at the center of the Brillouin zone (BZ) and ellipsoidal electron pockets at its corners. The experimental Fermi surface (FS) map in Fig.~1(c) was measured in the PM phase using 21.2 eV light from He-lamp. It is in good agreement with DFT calculations. The observed broad FS features are the result of a strong three-dimensionality of the band structure that is projected onto $k_z$, $k_y$ plane and limited $k_z$-selectivity of the ARPES experiment.

\begin{figure}[t]
    \includegraphics[width=1\linewidth]{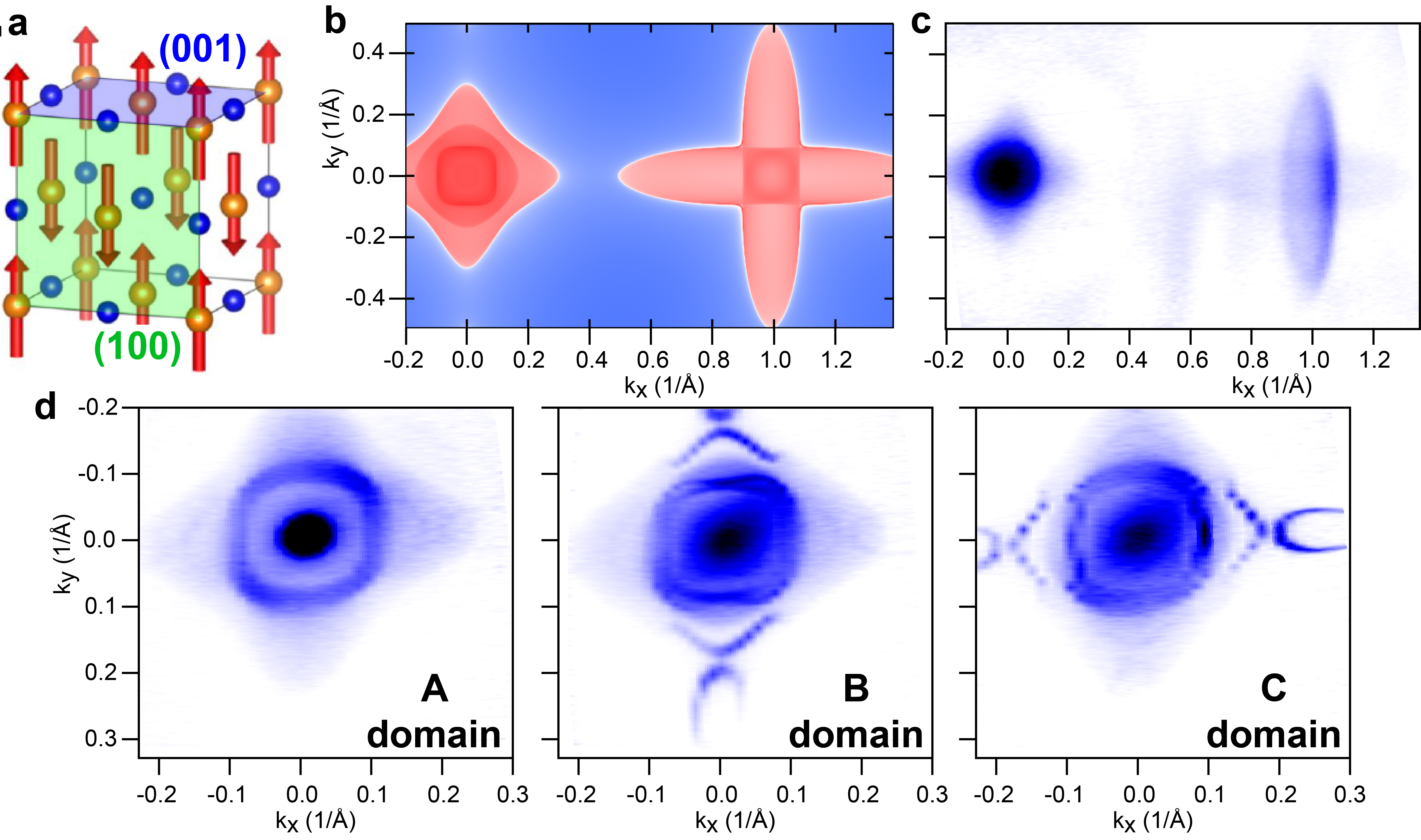}%
    \caption{NdSb crystal- and band-structure.
    (a) Schematic magnetic structure of AFM1 ordering. Green and blue squares represent (100/010) and (001) planes, respectively. 
    (b) DFT-calculated non-magnetic Fermi surface. 
    (c) Fermi surface map measured in the paramagnetic state (T~=~17.5~K). 
    (d) Three Fermi surface maps measured at different locations on one sample in the AFM state (T~=~5.5~K).}
\end{figure}

For further analysis, we measured detailed datasets around the center of the BZ using laser-based ARPES. Fig.~1(d) shows three FS maps measured at $T < T_N$ at different locations on the same sample surface cleave. All these maps differ from the map measured in the PM state (Fig.~1(c)) and from each other. 
As it will be shown later, the first map in Fig.~1(d) corresponds to the case of AFM1 ordering with magnetic moments oriented orthogonal to the sample surface (001 surface in Fig.~1(a)), while two other maps correspond to the case of AFM1 ordering with magnetic moments oriented parallel to the sample surface (100/010 surface in Fig.~1(a)) along the vertical and horizontal directions.

In Fig.~2, we analyze the effects of magnetic ordering on the electronic band structure at (001) surface. Parts of the DFT calculated FS near the Brillouin zone center for AFM1 (001), and PM phases are shown in Fig.~2(a) and (b), respectively. In the magnetically ordered state, we observe the suppression of the inner bulk pocket and the appearance of two sharp contours. This suppression is a result of the opening of a hybridization gap (for details, see Appendix B). The inner contour is formed by electron-like surface-state dispersion (see Fig.~2(c)), and the other contour is formed by a folded band and W-shaped surface-state dispersion. To analyze the development of these new features and compare them with the calculations, we measured datasets at several temperatures in the magnetically ordered and PM state from A-domain. In Fig.~2(d) and (e), we show the temperature evolution of the FS and the band dispersions along a high-symmetry direction. At $T =$~6.5K, the bottom of the electron-like and the W-shaped dispersion are observed in the experimental data as it is predicted by DFT calculations.

\begin{figure*}[t]
    \includegraphics[width=5.5 in]{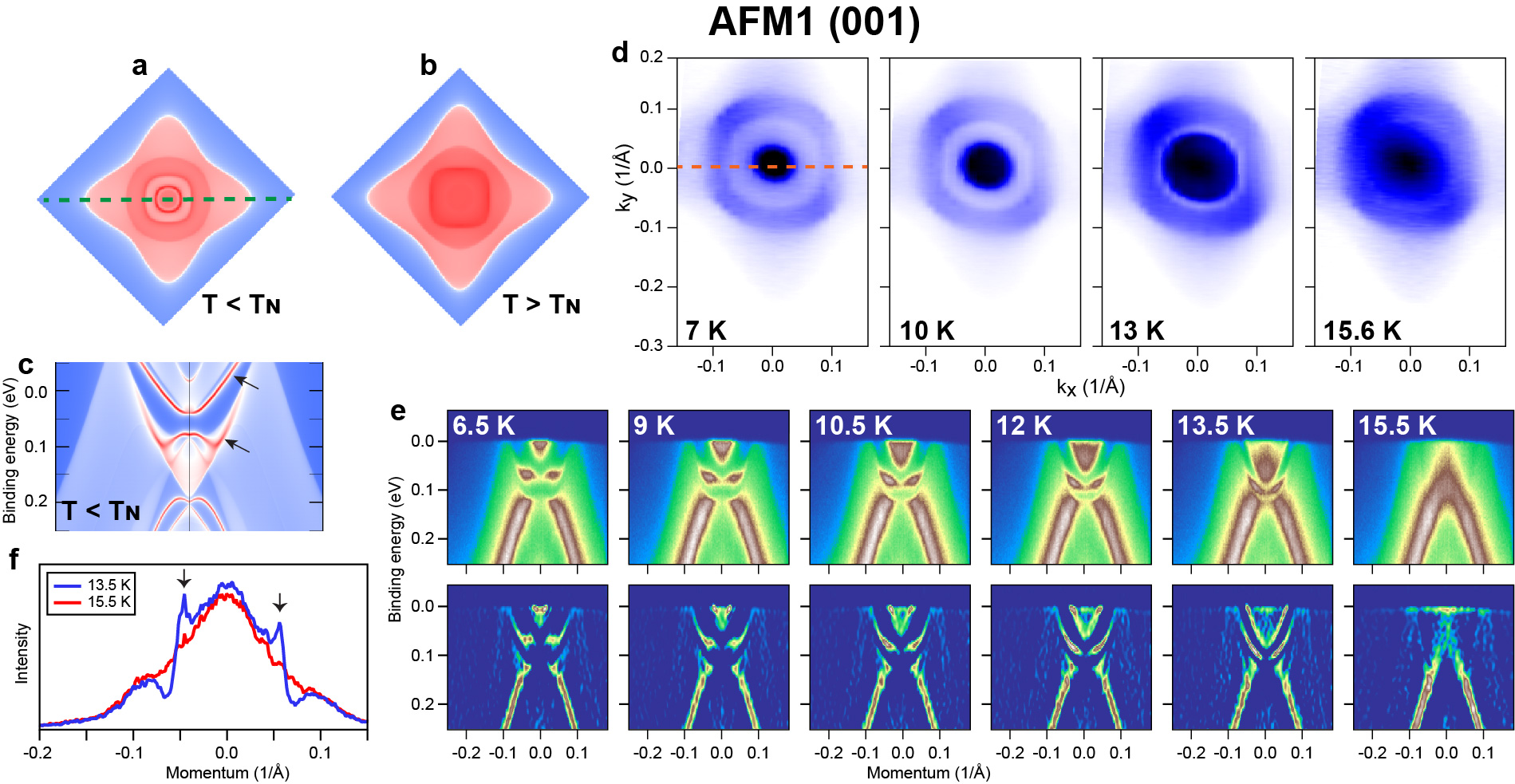}%
    \caption{Band structure at the (001) cleaving plane.
    (a) and (b) DFT-calculated part of the Fermi surface near $\Gamma$-point for AFM and PM phases of NdSb, respectively.
    (c) DFT-calculated band dispersions along the cut marked with a line in (a).
    (d) Temperature dependence of the Fermi surface from the A-domain.
    (e) Temperature dependence of band dispersions along the cut marked with a line in (d) and corresponding second derivative plots.
    (f) $E_F$ MDCs from spectra in (e). All the experimental data in this plot were measured using photon energy 1.67~eV.
    The arrows in (c) mark SS dispersions}
\end{figure*}

With the temperature increasing, both SS dispersions move to higher binding energies. At the same time, the W-shaped dispersion gradually changes its shape towards a regular parabola. As a result, the FS pockets increase in size. The inner FS pocket appears as a solid circle with a sharp perimeter. This happens because, besides the SS dispersion, which should form just a contour of the FS, there are bulk states above the SS parabola, as is seen from the DFT calculations in Fig.~2(c). In the experimental data, the SS parabola is the most pronounced in the 13.5 K spectrum near the Fermi level. The momentum distribution curve obtained from this spectrum at the Fermi energy ($E_F$) is shown in Fig.~2(f). Two sharp peaks associated with the surface states can be seen in this curve. Finally, in the paramagnetic state at T~=~15.5~K, both SS dispersions disappear. Also, from the plots in Fig.~2(e) and MDCs in Fig.~2(f), one can see that in the AFM1 state, a hybridization gap is opening inside the bulk states. These new SS are directly linked to this gap: the electron-like dispersion is the upper boundary of the gap, and the W-shaped dispersion exists inside the gap and partially merges with its lower boundary.

For further analysis, we measured a set of spectra along $\Gamma-X$ direction at T= 10.5 K using light with different photon energy. These spectra are shown in Fig.~3, together with the corresponding second-derivative plots. This allowed us to access parts of the BZ with different $k_z$ and distinguished SS from bulk states. Two SS dispersions mentioned above indeed do not shift with photon energy. The only difference between these spectra is the suppression of some parts of them at lower photon energy. This suppression is likely a result of matrix elements.
Interestingly, in 6.79 and 6.85 eV spectra, we observe a Dirac-like feature. However, the DFT (Fig.~3(c)) calculations do not predict a Dirac cone there. 
Two other hole-like SS dispersions predicted by DFT calculations can also be observed in the experiment. They correspond to the top of the band located at 300~meV and the another one  at 420~meV in the experimental data. We can also observe two more relatively sharp dispersions (marked with red arrows in Fig.~3). These dispersions shift with photon energy and are associated with bulk states.

\begin{figure*}[tb]
    \includegraphics[width=5.5 in]{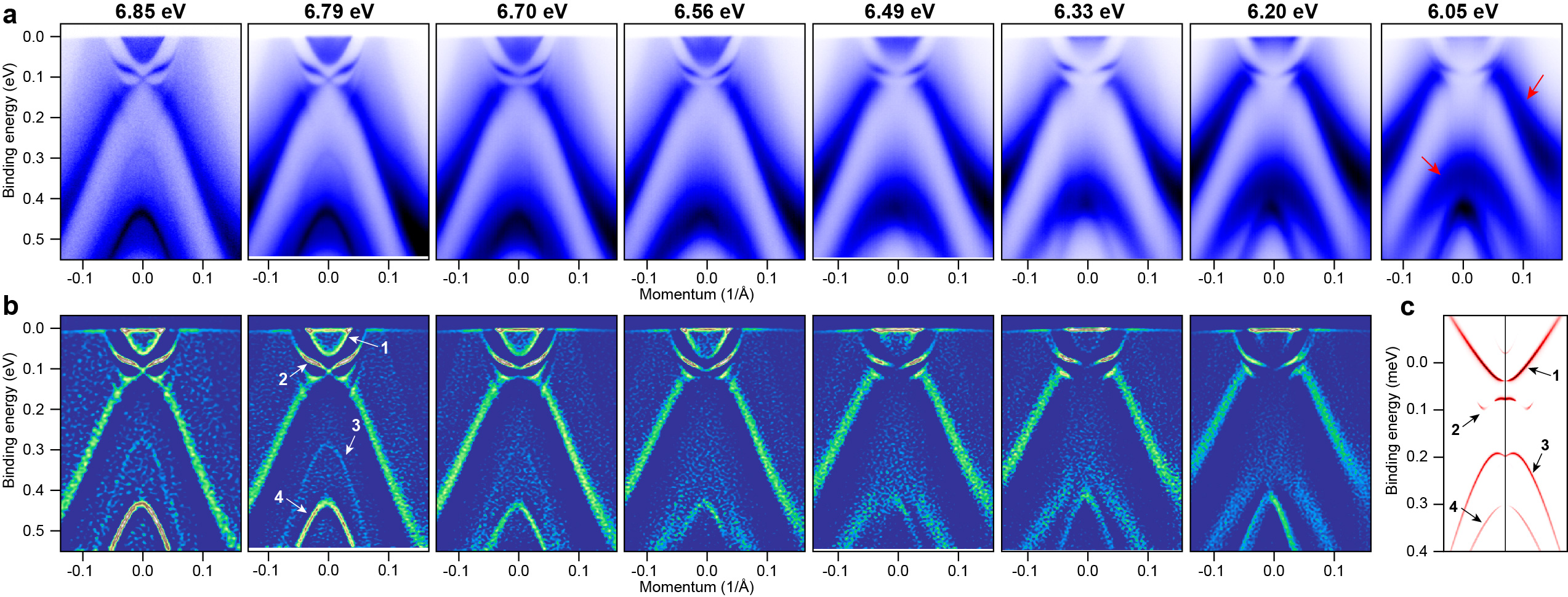}%
    \caption{Photon energy dependence.
    (a) Spectra measured along $\Gamma$-X direction using photon energy from 6.05 to 6.85~eV.
    (b) Corresponding plots of second derivatives for the spectra from (a).
    (c) DFT-calculated surface state dispersion.
    The arrows in (b) and (c) mark corresponding SS dispersions. The arrows in (a) mark bulk dispersions.}
\end{figure*}

In Fig.~4, we analyze the effects of (100/010) ordering on the surface state band dispersions. 
The DFT calculated FS and dispersions along the $\Gamma-X$ and $\Gamma-Y$ directions are shown in Fig.~4(a)-(c). These calculations predict the existence of several additional surface states $\Gamma-X$ in the magnetically ordered phase, but the most interesting for us are the two outermost surface states, which form Fermi arcs and elliptical electron pockets on the FS. Such states were already observed in the sibling compound NdBi \cite{SchrunkNature2022}. The calculations successfully reproduce the experimental FS from domains B and C, shown in Fig~1(d). Both maps demonstrate that the SS Fermi arcs and elliptical pockets are present only along one direction, within a given domain. For further analysis, we plot experimental dispersions along two orthogonal cuts for both maps in Fig.~4(d-g). The cuts along the $\Gamma-X$ direction show the presence of SS, namely pair of K-C split bands near the E$_f$ (panels d and f), which are responsible for the formation of Fermi arcs and elliptical pockets. At the same time, we observe no signs of the surface state near the $\Gamma$-point or suppression of the bulk states in this region similar to those observed in the A-domain data. The SS are seemingly absent along $\Gamma$-Y direction (Fig.~4(e) and (g)). Taking into account that datasets from all domains were measured under identical experimental conditions (experiment geometry, light polarization, and photon energy), we can conclude that the absence of SS states in the experimental data in one direction is an intrinsic property of this material and not caused by the matrix elements.

\begin{figure*}[t]
    \includegraphics[width=5.5 in]{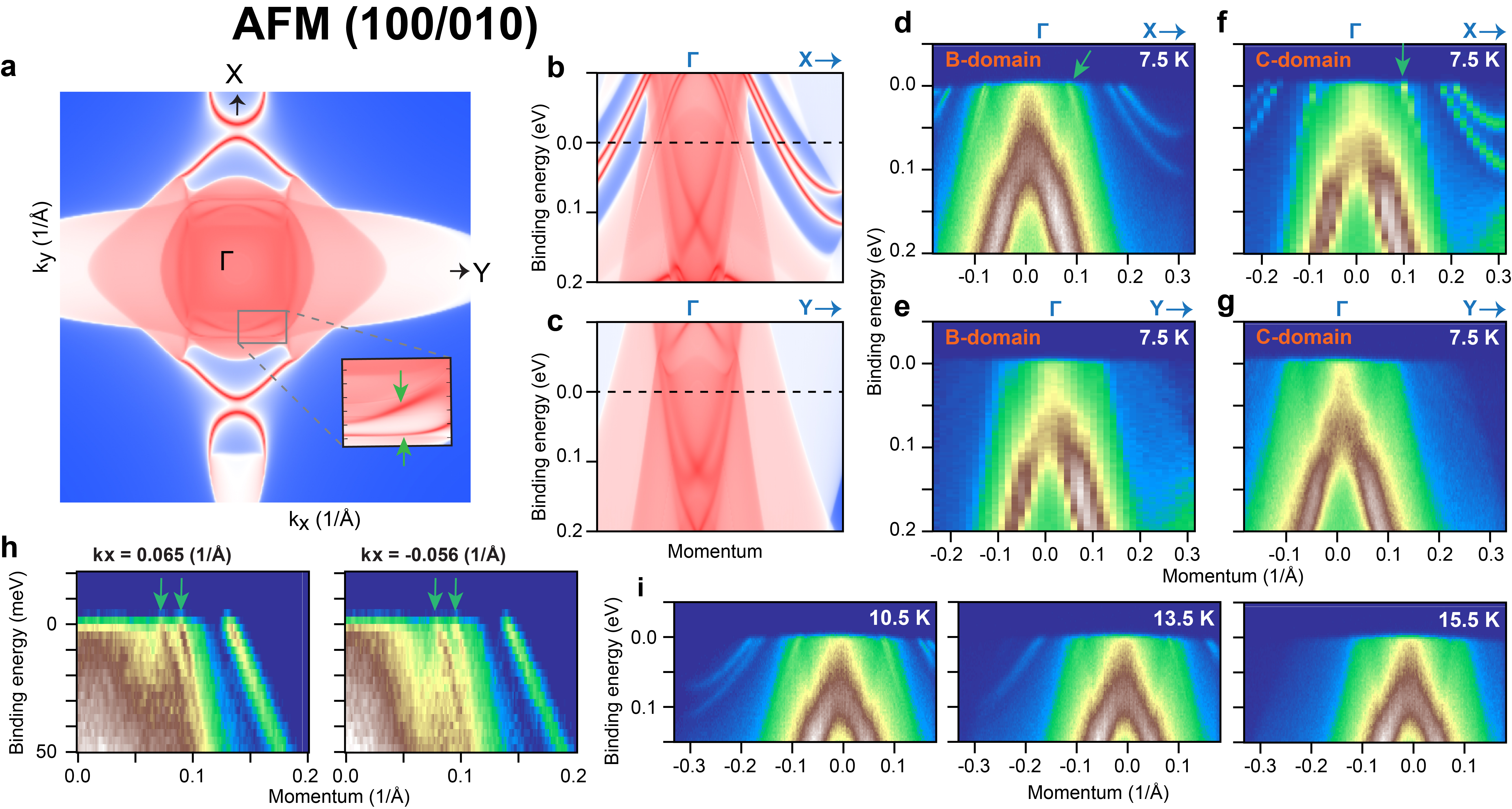}%
    \caption{Band structure in the case of (100/010) cleavage plane.
    (a) DFT-calculated Fermi surface for AFM1 state.
    (b) and (c) DFT-calculated surface state dispersions along $\Gamma-X$ and $\Gamma-Y$ in AFM1 state, respectively.
    (d) and (e) Experimental band structure along $\Gamma-X$ and $\Gamma-Y$ directions obtained from the B-domain dataset respectively.
    (f) and (g) The same as (d) and (e), respectively, but for the C-domain.
    (h) Cuts parallel to $\Gamma-X$ taken at $k_x =$ 0.065 $\AA^{-1}$ and -0.056 $\AA^{-1}$, respectively.
    (i) Temperature dependence of $\Gamma-X$ spectrum from (d).
    The arrows demonstrate the correspondence between particular SS features in the DFT calculations and the experimental data}
\end{figure*}

The temperature evolution of the spectrum from Fig.~4(d) is shown in Fig.~4(i). The splitting of the SS decreases with increased temperature, and the SS vanishes above $T_n$. This result agrees with the previous studies of NdSb, and NdBi \cite{kushnirenko2022rare, SchrunkNature2022}. This behavior is similar to the behavior of the SS on the (001) surface (Fig.~2(e)), where the distance between the surface state bands also decreases with increasing temperature. Moreover, the absolute positions of the bottoms of these bands are the same for both domain types. Their positions estimated for (100) surface at T~=~10.5 K (181 nm plot of Fig.~3) are 66$\pm$4~meV and 104$\pm$2~meV. Their positions estimated for (001) surface at T~=~10.5~K (Fig.~4(g)) are 68$\pm$2~meV and 105$\pm$2~meV. This similarity is not surprising since, in both cases, SS are formed inside the same hybridization gap.

In the paramagnetic state at T~$\sim$~15.5 K, SS dispersions disappear. The high-temperature spectra in Fig.~2(e) and Fig.~4(g) look identical, which indicates that both domain types are indeed formed in the same material. However, after several cycles of heating the sample to $T > T_N$ and cooling it down, we found the same domains at their original locations (for the details, see Appendix A). Thus, most likely, the domain type can be predetermined by some factors, e.g., pinned by the strain present in the crystal.

The spectra in Fig.~4(d) and (f) exhibit one more SS dispersion (marked with green arrows), which disappears in the paramagnetic state. Under closer inspection, one can see that this dispersion is split into two branches. This can be better seen in the cuts taken slightly off the high-symmetry direction where this splitting is large (see Fig.~4(h)). As well as the other SS dispersions, this splitting is predicted by DFT calculations. The corresponding features can be found in the calculated FS (Fig.4~(a)).

Elliptical SS pockets and arcs with four-fold symmetry were reported in NdBi in previous ARPES work \cite{SchrunkNature2022, kushnirenko2022rare}. In analogy to NdSb, these states can be associated with AFM1 (100/010) domains \cite{wang2022multi}. However, no features associated with (001) domains have been observed in NdBi.
In order to prove the presence of (001) domains, we performed more ARPES measurements on NdBi. The results of these  measurements are shown in Fig.~5. These spectra do not demonstrate the SS features observed in the previous studies, but they show a hybridization gap at the $\Gamma$ point and the development of several additional SS dispersions near the $\Gamma$ point. This makes this band structure similar to the band structure of the (001) domain of NdSb. Also, as well in NdSb, the additional SS, which are located in the gap, move up with the temperature decreasing. Thus, we can associate this band structure with AFM(001) ordered domain in NdBi.

\begin{figure}[b]
    \includegraphics[width=1\linewidth]{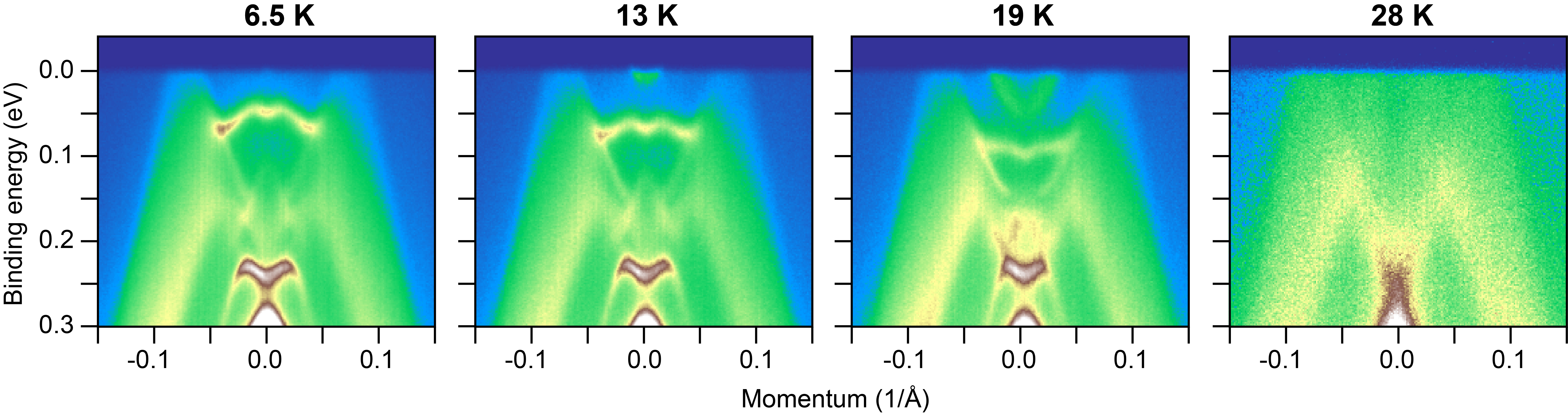}%
    \caption{AFM1 ordering in NdBi ((001) surface). Temperature dependence of band structure along $\Gamma-X$ direction. T$_N$~=~24~K}
\end{figure}

\section{Conclusions}
We investigated the evolution of the electron structure in NdSb and NdBi upon antiferromagnetic transition using ARPES and demonstrated the presence of domains with different configurations of surface states. This result agrees with the earlier neutron diffraction study \cite{mukamel1985phase}, which also reported the presence of three domain types in NdSb.

The development of a domain with magnetic moments oriented orthogonal to the sample surface causes the formation of a hybridization gap at the $\Gamma$ point and several additional surface-state dispersions near the $\Gamma$ point. 
The development of a domain with magnetic moments oriented parallel to the sample surface causes the formation of a different set of surface states, including an unconventional surface-state pair \cite{SchrunkNature2022} that forms an arc and an elliptical pocket on the Fermi surface. Except for a Dirac-like feature in (001) domain which remains unexplained, the observed electron structure of all three domains in NdSb is in exceptional agreement with our DFT calculations for AFM1 ordered phase.

\begin{acknowledgments}
This work was supported by the U.S. Department of Energy, Office of Basic Energy Sciences, Division of Materials Science and Engineering. Ames Laboratory is operated for the U.S. Department of Energy by Iowa State University under Contract No. DE-AC02-07CH11358. Crystal growth, characterization and were supported by Center for the Advancement of Topological Semimetals (CATS), an Energy Frontier Research Center funded by the U.S. DOE, Office of Basic Energy Sciences.
\end{acknowledgments}

\appendix

\begin{figure}[b]
    \includegraphics[width=1.0\linewidth]{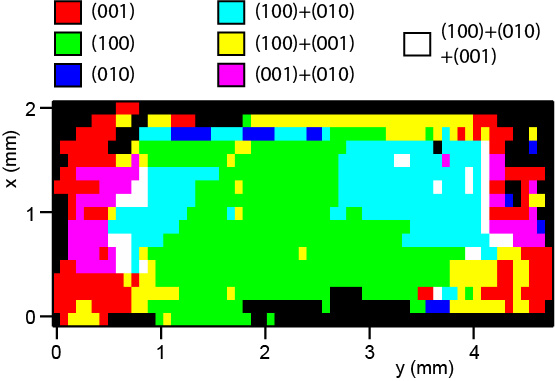}%
    \caption{NdSb: additional data. 
    (a) Domain map measured using ARPES. Red, green, and blue colors represent parts of the sample where one domain type. Cyan, magenta, yellow, and wight colors represent parts of the sample where we observed the superposition of signals from several domains.}
\end{figure}

\section{Reproducibility}
Most of the data presented in the main part of the manuscript were collected from one sample. The order in which data were collected is shown in Table I. We see that after heating the sample up and cooling it down, the band structure measured from the part of the sample where the B-domain was, did not change. Moreover, the band structure measured from the part of the sample where the A-domain was, did not change after three cycles of temperature variation. Also, it shows that under the experimental conditions, the aging of the sample surface is negligible: the map measured from the A-domain soon after cleaving the sample (Fig.~1(c)) looks identical to the similar map measured 24 hours later (Fig.~3(d)).

The mapping of another sample (Fig.~6(a)) has shown that it also has parts where each one of the three domain types is dominant, as well as parts where we observe the superposition of signals from several domains.

\begin{table}[h!]
\centering
\begin{tabular}{ |c|c|c|c|c| }
 \hline
 Time & Type & Domain & Temperature & Plot\\
 (hours) & & & & \\
 \hline
 0.9 & FS map & B & 5.7K & 1(c), 4(e,f)\\
 \hline
 2.6 & FS map & A & 5.7K & 1(c), 4(c,d)\\
 \hline
 4.3 & FS map & C & 5.7K & 1(c)\\
 \hline
 6.4 & spectrum & A & 6.5K & 2(e)\\
 \hline
 6.5 & spectrum & A & 9K & 2(e)\\
 \hline
 6.6 & spectrum & A & 10.5K & 2(e)\\
 \hline
 6.7 & spectrum & A & 12K & 2(e)\\
 \hline
 6.7 & spectrum & A & 13.5K & 2(e)\\
 \hline
 6.8 & spectrum & A & 15.5K & 2(e)\\
 \hline
 7.0 & spectrum & B & 10.5K & 4(i)\\
 \hline
 7.1 & spectrum & B & 13.5K & 4(i)\\
 \hline
 7.2 & spectrum & B & 15.5K & 4(i)\\
 \hline
 34.6 & FS map & A & 13K & 2(d)\\
 \hline
 36.4 & FS map & A & 11K & 2(d)\\
 \hline
 37.7 & FS map & A & 15.6K & 2(d)\\
 \hline
 38.5 & FS map & A & 7K & 2(d)\\
 \hline
\end{tabular}
 \caption{The order in which data were collected.}
\end{table}

\section{DFT calculations}
All density functional theory \cite{Hohenberg1964Inhomogeneous, Kohn1965Self} (DFT) calculations with spin-orbit coupling (SOC) were performed with the PBE \cite{Perdew1996Generalized} exchange-correlation functional using a plane-wave basis set and projector augmented wave method \cite{Blochl1994Projector}, as implemented in the Vienna Ab-initio Simulation Package (VASP) \cite{Kresse1996Efficient,Kresse1996Efficiency}. Using maximally localized Wannier functions \cite{Marzari1997Maximally,Souza2001Maximally}, tight-binding models were constructed to reproduce closely the band structure including SOC within $E_F\pm$1eV with Nd s-d-f and and Sb p orbitals. The surface spectral function and 2D Fermi surface (FS) were calculated with the surface Green’s function methods \cite{Sancho1984Quick,Sancho1985Highly} as implemented in WannierTools \cite{wu2018wanniertools}. In the DFT calculations, we used a kinetic energy cutoff of 254 eV, a Gaussian smearing of 0.05 eV, a $\Gamma$-centered Monkhorst-Pack \cite{Monkhorst1976Special} (11×11×8) k-point mesh for the 1q tetragonal unit cell and a (8×8×8) k-point mesh for the 2q cubic unit cell. For band structure calculations, we have used the experimental lattice parameters of  6.336 $\AA$. To account for the strongly localized Nd 4f orbitals in NdSb, an onsite Hubbard-like \cite{liechtenstein1995density} U=6.3 eV and J=0.7 eV have been used. Our DFT+U+SOC calculation on NdSb 1q and 2q gives a similar spin moment of 2.7 $\mu_B$ and an orbital moment of 5.7 $\mu_B$ in the opposite direction, resulting in a total magnetic moment of 3.0 $\mu_B$ on Nd.

\begin{figure}[b]
    \includegraphics[width=1\linewidth]{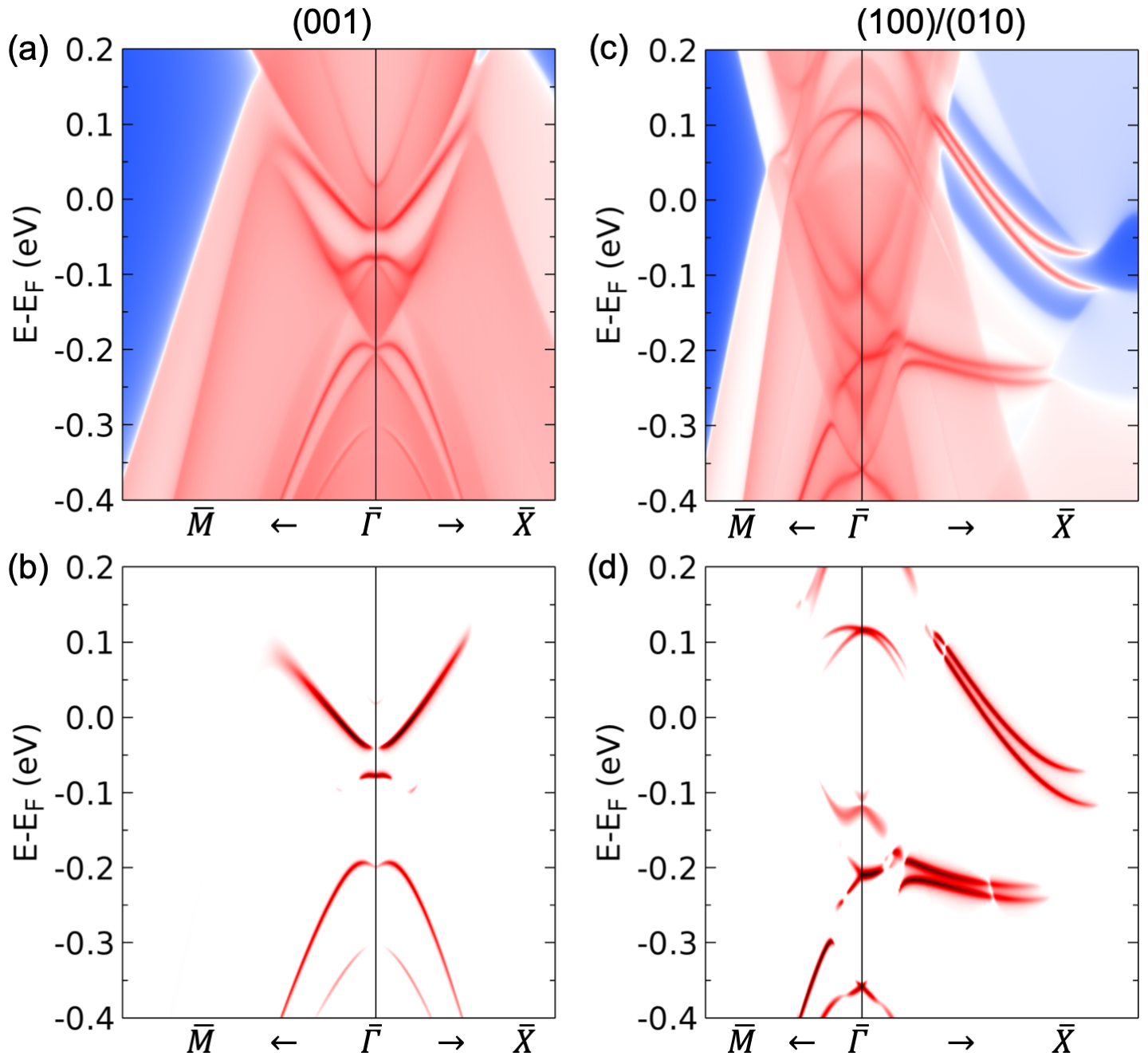}%
    \caption{DFT-calculated surface spectral function of NdSb AFM 1q on (a) (001) and (c) (100)/(010) surfaces. The corresponding surface only contributions are plotted in (b) and (d), respectively, to highlight the surface bands. Notably on (001), the hybridization gap and the associated surface states around the $\Gamma$ point as discussed in Fig.2 has an overall bulk band projection background, in contrast to the unconventional surface state pairs away from the $\Gamma$ point on (100)/(010) residing in the hybridization gap without such a background.}
\end{figure}

\bibliography{ndBi_arcs}

\end{document}